\documentclass[
    ,final            
  ]
  {aipproc}

\layoutstyle{8x11single}

\usepackage{graphicx}
\usepackage{latexsym}
\usepackage{amsfonts}
\usepackage{color}
\usepackage{colortbl}
\usepackage{amssymb,amsmath,enumerate,undertilde}
\usepackage{amsthm} 
\usepackage{bm}
\usepackage{indentfirst}
\usepackage{mathrsfs}
\usepackage[T1]{fontenc}
\usepackage{psfrag}

\newcommand{\Real}{\mathbb{R}}

\newcommand{\ve}[1]{\underline{#1}}

\newtheorem{lemma}{Lemma}

\begin{document}

\title{The evolution of a spatially homogeneous and isotropic universe filled with a collisionless gas}

\begin{abstract}
We review the evolution of a spatially homogeneous and isotropic universe described by a Friedmann-Robertson-Walker spacetime filled with a collisionless, neutral, simple, massive gas. The gas is described by a one-particle distribution function which satisfies the Liouville equation and is assumed to be homogeneous and isotropic. Making use of the isometries of the spacetime, we define precisely the homogeneity and isotropicity property of the distribution function, and based on this definition we give a concise derivation of the most general family of such distribution functions. For this family, we construct the particle current density and the stress-energy tensor and consider the coupled Einstein-Liouville system of equations. We find that as long as the distribution function is collisionless, homogenous and isotropic, the evolution of a Friedmann-Robertson-Walker universe exhibits a singular origin. Its future development depends upon the curvature of the spatial sections: spatially flat or hyperboloid universes expand forever and this expansion dilutes the energy density and pressure of the gas, while a universe with compact spherical sections reaches a maximal expansion, after which it reverses its motion and recollapses to a final crunch singularity where the energy density and isotropic pressure diverge. Finally, we analyze the evolution of the universe filled with the collisionless gas once a cosmological constant is included.
\end{abstract}


\author{Francisco Astorga}{
address={Instituto de F\'{\i}sica y Matem\'aticas,
Universidad Michoacana de San Nicol\'as de Hidalgo\\
Edificio C-3, Ciudad Universitaria, 58040 Morelia, Michoac\'an, M\'exico.}
}

\author{Olivier Sarbach}{
address={Instituto de F\'{\i}sica y Matem\'aticas,
Universidad Michoacana de San Nicol\'as de Hidalgo\\
Edificio C-3, Ciudad Universitaria, 58040 Morelia, Michoac\'an, M\'exico.},
altaddress={Gravitational Physics, Faculty of Physics, University of Vienna, Boltzmanngasse 5, 1090 Vienna, Austria.}
}

\author{Thomas Zannias}{
address={Instituto de F\'{\i}sica y Matem\'aticas,
Universidad Michoacana de San Nicol\'as de Hidalgo\\
Edificio C-3, Ciudad Universitaria, 58040 Morelia, Michoac\'an, M\'exico.}
}

\maketitle

\section{Introduction}

It has been recognized long ago that the relativistic kinetic theory of gases plays a vital role in the description of important processes in relativistic astrophysics and cosmology. Although the phenomenological description of matter by fluids or magneto-fluids seems to be  adequate for many astrophysical or cosmological scenarios, it fails to provide a reliable description for some important cases. For instance, the interaction of electromagnetic radiation with a a tenuous atmosphere requires both media to be treated by methods of kinetic theory. Similarly, during a core-collapse supernova explosion, the propagation of neutrinos in regions where their mean free path becomes comparable or larger than the core dimensions, methods of relativistic kinetic theory are required. In cosmology, the general relativistic Boltzmann equation is employed for the description of photons, neutrinos and often electrons during or before the recombination era. Calculations of the cosmological helium production treats the participating species of elementary particles via distribution functions satisfying the relativistic Boltzmann equation. Nowadays relativistic kinetic theory has become a respectable, challenging field in mathematical relativity. The Einstein-Liouville and the Einstein-Boltzmann equations are relevant and on the frontier of studies in mathematical relativity, see for instance Refs.~\cite{dByC73,aR04,gRaR92,mDaR07,nNmT09,Ringstrom-Book} and Ref.~\cite{hA11} for a recent review.

Early studies of relativistic kinetic theory started immediately after the birth of special relativity. Work by Synge~\cite{jS34}, Taubes and Weinberg~\cite{gTjW61}, Israel~\cite{wI63}, Lindquist~\cite{rL66}, Ehlers~\cite{jE71,jE73} and others led to the foundations for the modern description of the theory. In~\cite{oStZ13,oStZ14b,oStZ14a}, inspired by these early approaches, we presented an introduction to relativistic kinetic theory that relies on the geometric and symplectic properties of the tangent bundle of the background spacetime manifold. In~\cite{oStZ13}, the Hamiltonian structure of the theory has been highlighted, while in~\cite{oStZ14b} the importance of the Sasaki metric on the tangent bundle for the physical interpretation of the theory has been emphasized. As we have shown, the Sasaki metric is particularly useful to introduce a naturally defined integration theory on the tangent bundle which is free of ambiguities, and it is also useful in order to interpret symmetries of the distribution function.

Making use of these additional insights in the structure of the theory, in~\cite{oStZ14b} we also derived the most general spherically symmetric distribution function on an arbitrary spherically symmetric spacetime and the most general collisionless distribution function on a Kerr black hole background. In~\cite{oStZ14a} we extended these results to the derivation of a collisionless distribution function describing a charged gas on a Kerr-Newman black hole background.

While our earlier work focused on the formal structure of relativistic kinetic theory and astrophysical applications, in this work we discuss the propagation of a simple, collisionless gas from a cosmological perspective. At first, we discuss the properties of a neutral, simple, massive gas (that is, a collection of neutral, spinless classical particles of the same positive rest mass) propagating on a spatially homogenous and isotropic Friedmann-Robertson-Walker (FRW) spacetime,  and we derive the most general distribution function which shares the same symmetries as those of the FRW spacetime metric. As it turns out, such a distribution function has a very simple and tractable form: it is described by a function of proper time $\tau$ of the isotropic observers and a particular quantity $C$ which is constructed from the integrals of motion associated to the Killing vector fields of the spacetime. For the particular case of a \emph{collisionless}, spatially homogeneous and isotropic gas, the distribution function is required to satisfy the Liouville equation, and in this case it reduces to a function of the single quantity $C$, in accordance to previous results by Maharaj and Maartens~\cite{rMsM87a,rMsM87b,rMsM87c}. For early applications of relativistic kinetic theory to cosmology see also~\cite{jEpGrS68,lB69,eAlBjG75,jRjZ77}. In particular, in Ref.~\cite{eAlBjG75}, two relativistic collisionless gases are employed with the first one describing the population of normal galaxies and the other one the population of quasi-stellar objects. Their distribution function is chosen such that it maximizes an entropy functional, and the phenomenology of the model is analyzed in detail.

In a next step, we discuss the particle current density and stress-energy tensor of the gas. As long as the gas is described by a spatially homogeneous and isotropic distribution function, the particle current is parallel to the four-velocity of the isotropic observers while the stress-energy tensor has the form of a perfect fluid whose four-velocity (the dynamical mean velocity of the gas) is parallel to the particle current. However, the corresponding energy density $\rho$ and isotropic pressure $P$ are not arbitrary. Rather, they are given by fiber integrals of the distribution function over the future mass hyperboloid. Both, the density and pressure, are positive definite and the stress-energy tensor satisfies the weak, strong and dominant energy conditions. A particular interesting feature of the gas is that it behaves like a radiation fluid ($P\simeq \rho/3$) in the early universe, when the scale factor is very small, while in the late universe it behaves similarly to dust ($P\simeq 0$)~\cite{Ringstrom-Book}.

In the last part of this article, we use the stress-energy tensor in order to describe the self-gravity of the gas through Einstein's field equations, and we discuss a few qualitative properties of the spatially homogeneous and isotropic solutions of the Einstein-Liouville system. We show that spatially flat or hyperboloid universes expand forever, diluting the energy density and pressure of the gas, while a universe with compact spherical sections reaches a maximal volume, after which it reverses its motion and recollapses to a final crunch singularity. 

Motivated by the recent results coming from the analysis of distant supernovae suggesting that the universe is currently in an accelerated state of expansion~\cite{aR98,sP99}, we also discuss the case where a positive cosmological constant $\Lambda > 0$ is included. For spatially flat or hyperboloid universes there is again an everlasting expansion, but in contrast to the previous case with $\Lambda = 0$ where the expansion is damped due to the gravitational pull, now there is an accelerated expansion in the final stage of the evolution. For compact spherical sections the universe may expand forever or recollapse, depending on the values of the cosmological parameters.

Limitations of our approach and future work are discussed in the conclusions.

\section{Spatially homogeneous and isotropic collisionless distribution functions}

In this section, we consider a fixed FRW spacetime, and we review its symmetry properties and the derivation of the most general distribution function sharing these symmetries. Our derivation is based on the recent work in~\cite{oStZ14b}, where a systematic discussion on symmetries of distribution functions is provided. More specifically, it has been shown that infinitesimal generators $\xi$ of one-parameter groups of isometries of the base manifold $(M,g)$ lift naturally to infinitesimal generators $\hat{\xi}$ of one-parameter groups of isometries on the tangent bundle $(TM,\hat{g})$, where $\hat{g}$ is the Sasaki metric which is naturally induced from the spacetime metric $g$. Furthermore, it has been shown in~\cite{oStZ14b} that the lifted generator $\hat{\xi}$ is tangent to the mass shells $\Gamma_m := \{ (x,p)\in TM : g_x(p,p) = -m^2 \}$ associated to particles of rest mass $m$ and that it commutes with the Liouville vector field $L$: $[L,\hat{\xi}] = 0$. A distribution function $f$ on a particular mass shell $\Gamma_m$ is said to be invariant with respect to the isometry generated by $\hat{\xi}$ if the Lie derivative of $f$ with respect to $\hat{\xi}$ vanishes, $\pounds_{\hat{\xi}} f = 0$. Accordingly, if the spacetime manifold admits a Lie group $G$ of isometries, generated by Killing vector fields $\xi_1,\xi_2,\ldots,\xi_r$, the distribution function is said to be $G$-symmetric if $\pounds_{\hat{\xi}_a} f = 0$, $a=1,2,\ldots,r$.

\subsection{Killing algebra of the FRW spacetime}

In the case under consideration in this article the spacetime $(M,g)$ has the form $M = (0,\infty)\times \Sigma_k$ and
\begin{equation}
g = -d\tau^2 + a(\tau)^2\frac{dx^2 + dy^2 + dz^2}{\left( 1 + \frac{k}{4}|\ve{x}|^2 \right)^2},
\qquad
|\ve{x}| := \sqrt{x^2 + y^2 + z^2},
\label{Eq:FRWMetric}
\end{equation}
where $a(\tau)$ is the scale factor, and $\Sigma_k$ is either hyperbolic space $H^3$, Euclidean space $E^3$ or spherical space $S^3$ depending on the (normalized) value of the constant spatial curvature $k=-1$, $0$ or $1$. The metric~(\ref{Eq:FRWMetric}) is spatially homogeneous and isotropic, and the symmetry group $G$ is generated by the following particular Killing vector fields of $(M,g)$:
\begin{eqnarray}
\xi(\ve{s}) &:=& (\ve{s}\wedge \ve{x})\cdot\frac{\partial}{\partial\ve{x}},
\label{Eq:xi}\\
\eta(\ve{b}) &:=& \left( 1 - \frac{k}{4}|\ve{x}|^2 \right)\ve{b}\cdot\frac{\partial}{\partial\ve{x}}
 + \frac{k}{2}(\ve{b}\cdot\ve{x})\ve{x}\cdot\frac{\partial}{\partial\ve{x}},
\label{Eq:eta}
\end{eqnarray}
where $\ve{s},\ve{b}\in\Real^3$ are constant vectors, $\ve{x} := (x,y,z)$, and the wedge and the dot denote the standard vector and scalar products in $\Real^3$. The Killing vector fields $\xi(\ve{s})$ and $\eta(\ve{b})$ satisfy the following commutation relations for all $\ve{s}_1,\ve{s}_2,\ve{b}_1,\ve{b}_2\in\Real^3$:
\begin{eqnarray}
\left[ \xi(\ve{s}_1),\xi(\ve{s}_2) \right] &=& -\xi(\ve{s}_1\wedge \ve{s}_2),
\label{Eq:ComRel1}\\
\left[ \xi(\ve{s}_1),\eta(\ve{b}_2) \right] &=& -\eta(\ve{s}_1\wedge \ve{b}_2),
\label{Eq:ComRel2}\\
\left[ \eta(\ve{b}_1),\eta(\ve{b}_2) \right] &=& -k\xi(\ve{b}_1\wedge \ve{b}_2).
\label{Eq:ComRel3}
\end{eqnarray}
In particular, when $\ve{s}$ or $\ve{b}$ is equal to one of the basis vectors $\ve{e}_1 = (1,0,0)$, $\ve{e}_2 = (0,1,0)$ or $\ve{e}_3 = (0,0,1)$ it is sometimes convenient to introduce the infinitesimal generators $\xi_j := \xi(\ve{e}_j)$, $\eta_j := \eta(\ve{e}_j)$, $j = 1,2,3$ which are given explicitly by\footnote{The Killing vector fields $\eta_a$ correspond to the vector fields ${\bf X}_a$, $a=1,2,3$, in Ref.~\cite{rMsM87b}, and the vector fields $\xi_a$ to the vector fields ${\bf X}_{3+a}$, $a=1,2,3$, in that reference. However, notice that while the expressions in terms of Cartesian coordinates agree precisely with those of Eq.~(2) in~\cite{rMsM87b}, it seems there are misprints in the expressions for ${\bf X}_3$ and ${\bf X}_4$ in spherical coordinates in Eq.~(3) of that reference and in Eq.~(2) of Ref.~\cite{rMsM87c}: the expression for ${\bf X}_4$ is off by a minus factor and the last term in the expression for ${\bf X}_3$ should have a derivative with respect to $\theta$ instead of $\phi$.} 

\begin{eqnarray}
\xi_1 &=& y\frac{\partial}{\partial z} - z\frac{\partial}{\partial y}
 = -\sin\varphi\frac{\partial}{\partial\vartheta} 
 - \cot\vartheta\cos\varphi\frac{\partial}{\partial\varphi},\\
\xi_2 &=& z\frac{\partial}{\partial x} - x\frac{\partial}{\partial z}
 = \cos\varphi\frac{\partial}{\partial\vartheta} 
 - \cot\vartheta\sin\varphi\frac{\partial}{\partial\varphi},\\ 
\xi_3 &=& x\frac{\partial}{\partial y} - y\frac{\partial}{\partial x}
 = \frac{\partial}{\partial\varphi},
\end{eqnarray}
and
\begin{eqnarray}
\eta_1 &=& 
\left( 1 + \frac{k}{4}x^2 - \frac{k}{4}y^2 - \frac{k}{4}z^2 \right)\frac{\partial}{\partial x}
 + \frac{k}{2} x y\frac{\partial}{\partial y} + \frac{k}{2} x z\frac{\partial}{\partial z}
\nonumber\\
 &=& \left( 1 + \frac{k}{4} r^2 \right)\sin\vartheta\cos\varphi\frac{\partial}{\partial r}
 + \frac{1}{r}\left( 1 - \frac{k}{4} r^2 \right)
 \left( \cos\vartheta\cos\varphi\frac{\partial}{\partial\vartheta}  
  - \frac{\sin\varphi}{\sin\vartheta}\frac{\partial}{\partial\varphi} \right),\\
\eta_2 &=& \frac{k}{2} x y\frac{\partial}{\partial x}
 + \left( 1 - \frac{k}{4}x^2 + \frac{k}{4}y^2 - \frac{k}{4}z^2 \right)\frac{\partial}{\partial y}
 + \frac{k}{2} y z\frac{\partial}{\partial z}
\nonumber\\
 &=& \left( 1 + \frac{k}{4} r^2 \right)\sin\vartheta\sin\varphi\frac{\partial}{\partial r}
 + \frac{1}{r}\left( 1 - \frac{k}{4} r^2 \right)
 \left( \cos\vartheta\sin\varphi\frac{\partial}{\partial\vartheta}  
  + \frac{\cos\varphi}{\sin\vartheta}\frac{\partial}{\partial\varphi} \right),\\
\eta_3 &=& \frac{k}{2} x z\frac{\partial}{\partial x} + \frac{k}{2} y z\frac{\partial}{\partial y}
 + \left( 1 - \frac{k}{4}x^2 - \frac{k}{4}y^2 + \frac{k}{4}z^2 \right)\frac{\partial}{\partial z}
\nonumber\\
 &=& \left( 1 + \frac{k}{4} r^2 \right)\cos\vartheta\frac{\partial}{\partial r}
 - \frac{1}{r}\left( 1 - \frac{k}{4} r^2 \right)\sin\vartheta\frac{\partial}{\partial\vartheta},
\end{eqnarray}
where we have also given the expressions in terms of spherical coordinates $(r,\vartheta,\varphi)$ such that $\ve{x} = (x,y,z) = r(\sin\vartheta\cos\varphi,\sin\vartheta\sin\varphi,\cos\vartheta)$. The commutation relations~(\ref{Eq:ComRel1},\ref{Eq:ComRel2},\ref{Eq:ComRel3}) imply:
\begin{equation}
[\xi_1,\xi_2] = -\xi_3,\qquad
[\xi_1,\eta_2] = -\eta_3,\qquad
[\eta_1,\eta_2] = -k\xi_3,
\label{Eq:ComRelBis}
\end{equation}
and cyclic permutations. Note that for the spatially flat case, $k=0$, $\xi_j$ and $\eta_j$ are the infinitesimal generators of rotations and translations, respectively. When $k\neq 0$ the space $\Sigma_k$ is still spherically symmetric, so that the generators $\xi_j$ are the same as for $k=0$, but in this case the $\eta_j$'s cannot be interpreted as generators of translations.

\subsection{Most general homogeneous and isotropic distribution function on a FRW spacetime}

In order to derive the most general distribution function on a FRW spacetime which is spatially homogeneous and isotropic, we compute the lifted vector fields $\hat{\xi}(\ve{s})$ and $\hat{\eta}(\ve{b})$. In terms of adapted local coordinates $(x^\mu,p^\mu)$ on the tangent bundle $TM$, the lift $\hat{\xi}$ of a Killing vector field $\xi$ is given by (see, for instance~\cite{oStZ14b})
\begin{equation}
\hat{\xi} = \xi^\mu\frac{\partial}{\partial x^\mu} 
 + p^\alpha\frac{\partial\xi^\mu}{\partial x^\alpha}\frac{\partial}{\partial p^\mu}.
\end{equation}
For the lift of the Killing vector fields defined in Eqs.~(\ref{Eq:xi},\ref{Eq:eta}) this yields the following expressions:
\begin{eqnarray}
\hat{\xi}(\ve{s}) &=& (\ve{s}\wedge \ve{x})\cdot\frac{\partial}{\partial\ve{x}}
 + (\ve{s}\wedge \ve{p})\cdot\frac{\partial}{\partial\ve{p}},\\
\hat{\eta}(\ve{b}) &=& 
\left( 1 - \frac{k}{4}|\ve{x}|^2 \right)\ve{b}\cdot\frac{\partial}{\partial\ve{x}}
 + \frac{k}{2}\left[ (\ve{b}\cdot\ve{x})
 \left( \ve{x}\cdot\frac{\partial}{\partial\ve{x}} + \ve{p}\cdot\frac{\partial}{\partial\ve{p}} \right)
 + (\ve{b}\cdot\ve{p})\ve{x}\cdot\frac{\partial}{\partial\ve{p}} 
 - (\ve{x}\cdot\ve{p})\ve{b}\cdot\frac{\partial}{\partial\ve{p}} \right],
\end{eqnarray}
where here and below, $\ve{p} := (p^x,p^y,p^z)$.

Now consider a smooth function $f: TM\to\Real$ on the tangent bundle $TM$ which is $G$-symmetric, that is, invariant with respect to the flows generated by $\hat{\xi}(\ve{s})$ and $\hat{\eta}(\ve{b})$ for all $\ve{s},\ve{b}\in\Real^3$, and let us characterize such functions. We start with the simpler case $k=0$ of a spatially flat universe. As commented above, in this case the Killing vector fields $\eta(\ve{b})$ generate translations in directions of $\ve{b}$, and we have simply
$$
\hat{\eta}(\ve{b}) = \ve{b}\cdot\frac{\partial}{\partial\ve{x}}.
$$
Therefore, invariance with respect to these translations implies that the distribution function must have the form
\begin{equation}
f = f(\tau,p^\tau,p^x,p^y,p^z),
\label{Eq:Homogeneousf}
\end{equation}
that is, independent of $\ve{x}$. Then, invariance with respect to rotations implies that
$$
(\ve{s}\wedge\ve{p})\cdot\frac{\partial f}{\partial\ve{p}} = 0,
$$
for all $\ve{s}\in\Real^3$ which, in turn, implies that
$$
f(x,p) = F(\tau,p^\tau,|\ve{p}|),\qquad |\ve{p}| := \sqrt{(p^x)^2 + (p^y)^2 + (p^z)^2},
$$
as can be shown by introducing spherical coordinates for $\ve{p}$, for instance.

When $k\neq 0$ the vector fields $\hat{\eta}(\ve{b})$ on the tangent bundle $TM$ are more complicated. However, it is not difficult to deduce the form of the invariant distribution function after the following two observations. First, it is convenient to replace the adapted local coordinates $(x^\mu,p^\mu)$ used so far with new local coordinates $(x^\mu,p^{a'})$ on $TM$, where $p^{a'}$ refer to the components of $p$ with respect to a local orthonormal frame. Specifically, we define
$$
p^{0'} := p^\tau,\qquad
\ve{p}' := \frac{a(\tau)}{1 + \frac{k}{4}|\ve{x}|^2}\ve{p},
$$
such that $g(p,p) = -(p^{0'})^2 + |\ve{p}'|^2$. With respect to these new coordinates the lifted generators assume the slightly simpler form
\begin{eqnarray}
\hat{\xi}(\ve{s}) &=& (\ve{s}\wedge \ve{x})\cdot\frac{\partial}{\partial\ve{x}}
 + (\ve{s}\wedge \ve{p}')\cdot\frac{\partial}{\partial\ve{p}'},\\
\hat{\eta}(\ve{b}) &=& 
\left( 1 - \frac{k}{4}|\ve{x}|^2 \right)\ve{b}\cdot\frac{\partial}{\partial\ve{x}}
 + \frac{k}{2}\left[ (\ve{b}\cdot\ve{x})\ve{x}\cdot\frac{\partial}{\partial\ve{x}} 
 + (\ve{b}\cdot\ve{p}')\ve{x}\cdot\frac{\partial}{\partial\ve{p}'} 
 - (\ve{x}\cdot\ve{p}')\ve{b}\cdot\frac{\partial}{\partial\ve{p}'} \right].
\end{eqnarray}
The second observation is based on the vector identity
$$
(\ve{b}\cdot\ve{p}')\ve{x} - (\ve{x}\cdot\ve{p'})\ve{b} 
 = (\ve{b}\wedge \ve{x})\wedge \ve{p}',
$$
which allows us to rewrite $\hat{\eta}(\ve{b})$ in the simpler form
\begin{equation}
\hat{\eta}(\ve{b}) = 
\left( 1 + \frac{k}{4}|\ve{x}|^2 \right)\ve{b}\cdot\frac{\partial}{\partial\ve{x}}
 + \frac{k}{2}\hat{\xi}(\ve{b}\wedge\ve{x}).
\end{equation}
Therefore, if $f$ is invariant with respect to the flows generated by $\hat{\xi}(\ve{s})$ and $\hat{\eta}(\ve{s})$, it follows that $\pounds_{\hat{\xi}(\ve{b}\wedge\ve{x})} f = 0$ and we conclude that $f$ must have the same form as in Eq.~(\ref{Eq:Homogeneousf}) except that $\ve{p} = (p^x,p^y,p^z)$ should be replaced by $\ve{p}'$. Then, $\pounds_{\hat{\xi}(\ve{s})} f = 0$ implies as before that $f$ can only depend on $\tau$, $p^{0'}$ and $|\ve{p}'|$.

We conclude that the most general spatially homogeneous and isotropic distribution function on a FRW spacetime has the form
\begin{equation}
f(x,p) = F\left( \tau,p^\tau,\frac{|\ve{p}|}{1 + \frac{k}{4}|\ve{x}|^2} \right),\qquad
|\ve{p}| := \sqrt{(p^x)^2 + (p^y)^2 + (p^z)^2},
\label{Eq:HomIsof}
\end{equation}
for some function $F$. If we only consider a simple gas of particles with the same positive rest mass $m > 0$, then it is sufficient to work on the associated future mass shell $\Gamma_m^+$, from which
$$
p^\tau = \sqrt{m^2 + a(\tau)^2\frac{|\ve{p}|^2}{\left( 1 + \frac{k}{4}|\ve{x}|^2 \right)^2} }
$$
can be eliminated.

\subsection{Most general collisionless, homogeneous and isotropic distribution function}

Our approach so far was merely based on symmetry considerations. Now we wish to derive the most general distribution function $f$ on a FRW spacetime which, in addition of being spatially homogeneous and isotropic, is also collisionless. This means that $f$ has to satisfy the Liouville (or collisionless Boltzmann) equation
\begin{equation}
\pounds_L f = p^\mu\frac{\partial f}{\partial x^\mu} 
 - \Gamma^\mu{}_{\alpha\beta} p^\alpha p^\beta\frac{\partial f}{\partial p^\mu} = 0.
\label{Eq:Liouville} 
\end{equation}

For the derivation of $f$, we first recall the following Lemma (see, for instance, Proposition 6 in Ref.~\cite{oStZ14b}):

\begin{lemma}
Let $\xi$ be the generator of a one-parameter group of isometries on $(M,g)$. Then, the quantity $Q_\xi(x,p) := g_x(p,\xi)$ is conserved along the flow generated by the Liouville vector field $L$.
\end{lemma}

As a consequence of this lemma, the six quantities $J_a := Q_{\xi_a} = g(p,\xi_a)$, $K_a := Q_{\eta_a} = g(p,\eta_a)$, $a = 1,2,3$, are conserved along the flow of $L$. For the following, we consider the important quantity
\begin{equation}
C^2 := |\ve{K}|^2 + k|\ve{J}|^2 = \sum\limits_{a=1,2,3}\left( K_a^2 + k J_a^2 \right),
\label{Eq:Casimir}
\end{equation}
which is also conserved along the Liouville flow. As the next lemma shows, it follows from the commutation relations (\ref{Eq:ComRel1},\ref{Eq:ComRel2},\ref{Eq:ComRel3}) that $C^2$ is also invariant with respect to the flows generated by the vector fields $\hat{\xi}(\ve{s})$ and $\hat{\eta}(\ve{b})$.

\begin{lemma}
We have $\pounds_{\hat{\xi}(\ve{s})} C^2 = \pounds_{\hat{\eta}(\ve{b})} C^2 = 0$ for all $\ve{s},\ve{b}\in\Real^3$.
\end{lemma}

\proof In order to show this we first note the following identity which is valid for any two Killing vector fields $\xi,\eta$ on $(M,g)$:
\begin{equation}
\pounds_{\hat{\xi}} Q_\eta = Q_{[\xi,\eta]},
\label{Eq:LittleId}
\end{equation}
where $Q_\eta(x,p) = g_x(p,\eta)$ is the conserved quantity associated to $\eta$. This identity follows almost immediately from the fact that the lifted generator $\hat{\xi}$ is a Killing vector field of $(TM,\hat{g})$ that commutes with the Liouville vector field $L$, see section 5 in Ref.~\cite{oStZ14b}. Using these observations we find
$$
\pounds_{\hat{\xi}} Q_\eta 
 = \pounds_{\hat{\xi}} [\hat{g}(\hat{\eta},L)]
 = \hat{g}(\pounds_{\hat{\xi}}\hat{\eta},L)
 = \hat{g}([\hat{\xi},\hat{\eta}],L)
 = g([\xi,\eta],p)
 = Q_{[\xi,\eta]},
$$
where we have also used Lemma 10 in Ref.~\cite{oStZ14b} in the last step.

Using the identity~(\ref{Eq:LittleId}) and the commutation relations~(\ref{Eq:ComRelBis}) it follows for all $i=1,2,3$:
\begin{eqnarray*}
\pounds_{\hat{\xi}_i} C^2 
&=& 2\sum\limits_{a=1,2,3} \left( K_a Q_{[\xi_i,\eta_a]} + k J_a Q_{[\xi_i,\xi_a]} \right)
 = -2\sum\limits_{a,b=1,2,3} \varepsilon_{iab}( K_a K_b + k J_a J_b ) = 0,\\
\pounds_{\hat{\eta}_i} C^2
&=& 2\sum\limits_{a=1,2,3} \left( K_a Q_{[\eta_i,\eta_a]} + k J_a Q_{[\eta_i,\xi_a]} \right)
 = -2\sum\limits_{a,b=1,2,3} \varepsilon_{iab}( k K_a J_b + k J_a K_b ) = 0,
\end{eqnarray*}
which proves the lemma.
\qed

Therefore, the quantity $C^2$ is invariant with respect to both the Liouville flow and the symmetry group. An explicit calculation based on the expressions~(\ref{Eq:xi},\ref{Eq:eta}) reveals that
\begin{equation}
C = a(\tau)^2\frac{|\ve{p}|}{1 + \frac{k}{4}|\ve{x}|^2},\qquad
|\ve{p}| := \sqrt{(p^x)^2 + (p^y)^2 + (p^z)^2}.
\label{Def:C}
\end{equation}
Comparing this with Eq.~(\ref{Eq:HomIsof}) we see that the most general spatially homogeneous and isotropic distribution function on a FRW spacetime can also be written as
\begin{equation}
f(x,p) = F(\tau,m,C),
\label{Eq:HomIsofBis}
\end{equation}
for some smooth function $F$. The previous Lemma provides an alternative explanation for the invariance of this $f$ under the symmetry group $G$: it is a function of the three quantities $\tau$, $m$ and $C$, all of which are invariant with respect to the flows generated by the lifted generators of the symmetry group. The arguments presented in the previous subsection show that this is the most general distribution function which is $G$-symmetric.

Introducing Eq.~(\ref{Eq:HomIsofBis}) into the Liouville equation, it is straightforward to derive from this the most general $G$-symmetric, collisionless distribution function: Since  $m$ and $C$ are conserved along the Liouville flow, the Liouville equation~(\ref{Eq:Liouville}) reduces simply to
$$
p^\tau\frac{\partial F}{\partial\tau} = 0,
$$
implying that $F$ is independent of $\tau$. Therefore, we arrive at the simple result that the most general \emph{collisionless}, spatially homogeneous and isotropic distribution function on a FRW spacetime must have the form
\begin{equation}
f(x,p) = F(m,C).
\label{Eq:HomIsoDistr}
\end{equation}

\section{Observables: current density and stress-energy tensor}

In this section, we evaluate the most important observables of the theory, namely the current density $J$ and the stress-energy tensor $T$ assuming a distribution function of the form $f(x,p) = F(m,C)$, see Eq.~(\ref{Eq:HomIsoDistr}). We only consider a simple gas, where each gas particle has the same rest mass $m > 0$. Therefore, since $m$ is fixed, we omit the explicit dependance of $F$ from $m$ in what follows. The observables $J$ and $T$ are defined as fiber integrals over the future mass hyperboloidal
$$
P_x^+ := \{ p\in T_x M : g_x(p,p) = -m^2, \hbox{$p$ future directed} \}
$$
with associated volume element $\pi_x$, and their geometrical invariant definition is given by (see~\cite{oStZ13,oStZ14b})
\begin{eqnarray}
J_x(\alpha) := \int\limits_{P_x^+} f(x,p) \alpha(p) \pi_x && \hbox{(current density)},
\label{Eq:CurrentDensity}\\
T_x(\alpha,\beta) := \int\limits_{P_x^+} f(x,p) \alpha(p)\beta(p) \pi_x 
&& \hbox{(stress-energy tensor)},
\label{Eq:StressEnergy}
\end{eqnarray}
where $\alpha,\beta\in T_x^* M$ are covectors at $x\in M$. Since $J_x$ is linear in $\alpha$ and $T_x$ is bilinear in $(\alpha,\beta)$, they define a vector field and a symmetric, contravariant tensor field on $M$, respectively. Their components relative to local coordinates $(x^\mu)$ of $M$ are obtained from $J^\mu(x) = J_x(dx^\mu)$, $T^{\mu\nu}(x) = T_x(dx^\mu,dx^\nu)$, which yields
\begin{eqnarray}
J^\mu(x) &=& \int\limits_{P_x^+} f(x,p) p^\mu \pi_x,
\label{Eq:CurrentDensityCoords}\\
T^{\mu\nu}(x) &=& \int\limits_{P_x^+} f(x,p) p^\mu p^\nu \pi_x.
\label{Eq:StressEnergyCoords}
\end{eqnarray}
As long as $f$ satisfies the Liouville equation, these observables are divergence-free, and if $f$ is nonnegative it follows that $J$ is future directed while $T$ satisfies the familiar energy conditions (for a proof of these assertions see, for instance, Ref.~\cite{oStZ13,oStZ14b}).

For a spatially homogeneous and isotropic  distribution function $f(x,p) = F(C)$ the right-hand sides of Eqs.~(\ref{Eq:CurrentDensityCoords},\ref {Eq:StressEnergyCoords}) can be simplified. For this we introduce an orthonormal basis in the tangent space $T_x M$, defined by
\begin{equation}
\left( e_{0},e_{x},e_{y}, e_{z} \right) := \left(
\frac{\partial}{\partial \tau}, 
\frac{1 + \frac{k}{4}|\ve{x}|^2}{a(\tau)}\frac{\partial}{\partial x},
\frac{1 + \frac{k}{4}|\ve{x}|^2}{a(\tau)}\frac{\partial}{\partial y},
\frac{1 + \frac{k}{4}|\ve{x}|^2}{a(\tau)}\frac{\partial}{\partial z}
\right), 
\label{Orth.basis}
\end{equation}
so that any vector $p\in P_x^+$ can be written in the form
$$
p = p^{\mu}\frac{\partial}{\partial x^{\mu}} 
 = p^{0'} e_{0} + p^{x'} e_{x} + p^{y'} e_{y} + p^{z'} e_{z}
$$
with $p^{0'} = \sqrt{m^2 + (p^{x'})^2 + (p^{y'})^2 + (p^{z'})^2}$. Passing to a spherical polar representation, we get the equivalent form     
\begin{equation}
p = \sqrt{m^2 + |\ve{p}'|^2} e_{0} + |\ve{p}'| n^{i} e_{i},
\label{Def.p}
\end{equation}
with $(n^x,n^y,n^z) = (\sin\theta\cos\phi,\sin\theta\sin\phi,\cos\theta)$. Relative to these frames, the Lorentz-invariant volume element $\pi_x$ takes the form
\begin{equation}
\pi_x = \frac{dp^{x'}\wedge dp^{y'} \wedge dp^{z'}}
{\sqrt{m^2 + (p^{x'})^2 + (p^{y'})^2 + (p^{z'})^2}}
 =\frac{|\ve{p}'|^2d|\ve{p'}|\wedge (\sin\theta d\theta )\wedge d\phi}{\sqrt{m^2 + |\ve{p}'|^2}},
\label{Lor.Meas}
\end{equation}
while for a spatially homogeneous and isotropic distribution function belonging to a simple gas we have
\begin{equation}
f(x,p) = F(C) = F(a(\tau)|\ve{p}'|), 
\label{DisFun}
\end{equation}
where the quantity $C$ has been defined in Eq.~(\ref{Def:C}), and we recall that here $|\ve{p}'|$ refers to the norm of the spatial components of $p\in P_x^+$ with respect to the \emph{orthonormal} basis defined in Eq.~(\ref{Orth.basis}), such that $C = a(\tau)|\ve{p}'|$.

Substituting Eqs.~(\ref{Lor.Meas},\ref{DisFun}) into Eq.~(\ref{Eq:CurrentDensityCoords}), respectively Eq.~(\ref{Eq:StressEnergyCoords}), we find that the frame components of the current density and the stress-energy tensor take the form 
\begin{eqnarray}
J^{\mu'}(x) &=& \int\limits_0^\infty\int\limits_0^\pi\int\limits_0^{2\pi} 
F(a(\tau)|\ve{p}'|) p^{\mu'} \frac{|\ve{p}'|^2\sin\theta d\phi d\theta d|\ve{p}'|}
{\sqrt{m^2 + |\ve{p}'|^2}},
\\
T^{\mu'\nu'}(x) &=& \int\limits_0^\infty\int\limits_0^\pi\int\limits_0^{2\pi} 
F(a(\tau)|\ve{p}'|) p^{\mu'} p^{\nu'} \frac{|\ve{p}'|^2\sin\theta d\phi d\theta d|\ve{p}'|}
{\sqrt{m^2 + |\ve{p}'|^2}}.
\end{eqnarray}
Using Eq.~(\ref{Def.p}) and changing the variable of integration from $|\ve{p}'|$ to $C = a(\tau)|\ve{p}'|$ we find that the current density is described by
\begin{equation}
J(x) = J^{\mu'}(x) e_{\mu'}
 = \left( \frac {4\pi}{a^{3}(\tau)} \int\limits_{0} ^{\infty} F(C) C^2 dC \right) e_{0},
\label{FramCom.Cur}
\end{equation}
i.e. $J$ is parallel to the four-velocity of the isotropic observers (those that see the universe homogenous and isotropic). This property of $J$ originates in the spatially homogenous and isotropic property of the distribution function. A similar computation shows that the nonvanishing frame components of the stress-energy tensor take the form:
\begin{eqnarray}
&& T^{0'0'} = \frac {4\pi}{ a^4(\tau)} \int\limits_0^\infty F(C) \sqrt{(ma(\tau))^2 + C^2}C^2dC,
\label{00Com.SE}\\
&& T^{x'x'} =T^{y'y'} =T^{z'z'} = \frac {4\pi}{3a^4(\tau)}
\int\limits_0^\infty F(C){\frac{C^4}{\sqrt{(ma(\tau))^2 + C^2}}} dC.
\label{SCom.SE}
\end{eqnarray}

Therefore, the current density and stress-energy tensor have the same form as the one of a perfect fluid, for which
$$
J^\mu = n(\tau) u^\mu,\qquad
T^{\mu\nu} = [\rho(\tau) + P(\tau)]u^{\mu}u^{\nu} + P(\tau)g^{\mu\nu}
$$
with $u = \frac{\partial}{\partial \tau}$ the dynamical mean velocity of the gas, $n(\tau)$ the particle density and $\rho(\tau)$ the energy density of the gas measured by observers comoving with the mean velocity, and $P(\tau)$ is the mean or isotropic pressure measured by such observers. In our case, the densities and pressure are given by
\begin{eqnarray}
n(\tau) &=& \frac {4\pi}{a^{3}(\tau)} \int\limits_{0} ^{\infty} F(C) C^2 dC,
\label{Eq:n}\\
\rho(\tau) &=& \frac{4\pi}{a^4(\tau)} \int\limits_0^\infty F(C)\sqrt{(ma(\tau))^2 + C^2} C^2 dC,\qquad 
P(\tau) = \frac {4\pi}{3a^4(\tau)} \int\limits_0^\infty 
F(C){\frac{C^4}{\sqrt{(ma(\tau))^2 + C^2}}} dC.
\label{Eq:rhoP}
\end{eqnarray}
Once the function $F(C)$ and the scale factor $a(\tau)$ have been specified, the current density and the stress-energy tensor of the gas are uniquely determined. 

The expressions in Eq.~(\ref{Eq:rhoP}) imply
$$
\rho(\tau) = 3P(\tau) + \frac{4\pi m^2}{a^2(\tau)} 
\int\limits_0^\infty F(C){\frac{C^2}{\sqrt{(ma(\tau))^2 + C^2}}} dC,
$$
and thus we always have $\rho \geq 3P \geq 0$ since the distribution function is nonnegative. These inequalities imply that the stress-energy tensor satisfies the weak, the strong and the dominant energy conditions, as is generally true for a kinetic gas (cf. Ref.~\cite{oStZ13}). It is also a simple matter to verify the equations $\nabla_\mu J^\mu = 0$, $\nabla_\mu T^{\mu\nu} = 0$, which in the spatially homogeneous and isotropic case reduce to
\begin{equation}
\frac{dn}{d\tau} + 3n\frac{d\log a}{d\tau} = 0,\qquad
\frac{d\rho}{d\tau} + 3(\rho + P)\frac{d\log a}{d\tau} = 0,
\label{Eq:DivTZero}
\end{equation}
from the explicit expressions given in Eqs.~(\ref{Eq:n},\ref{Eq:rhoP}).

Finally, we observe that in the limit where $ma\to 0$ which occurs either as $m\to 0$ (photon gas), or for fixed $m > 0$ as $a\to 0$, one gets the equation of state $P = \frac{1}{3}\rho$. Therefore, in these limits the gas behaves as a radiation fluid. Furthermore, expanding the integrals for the case in which the scale factor approaches zero it is seen that the energy density and pressure both diverge as $a^{-4}$, as for a radiation fluid. In the opposite limit, that is, when $a$ becomes very large, the energy density and pressure decay to zero as $a^{-3}$ and $a^{-5}$, respectively, and the collisionless kinetic gas behaves more and more like dust. In order to see this more explicitly, it is instructive to consider the particular case for which the distribution function is supported in a very narrow interval around some constant $C = C_0 > 0$:
\begin{equation}
F(C) = \frac{N}{4\pi C^2}\delta(C - C_0),
\label{Eq:DeltaF}
\end{equation}
for some dimensionless constant $N > 0$, where $\delta$ denotes the Dirac distribution. In this case, we obtain from Eqs.~(\ref{Eq:n},\ref{Eq:rhoP}),
\begin{equation}
n(\tau) = \frac{N}{a^3(\tau)},\qquad
\rho(\tau) = \frac{N}{a^4(\tau)}\sqrt{(ma(\tau))^2 + C_0^2},\qquad
P(\tau) = \frac{N}{3 a^4(\tau)}\frac{C_0^2}{\sqrt{(ma(\tau))^2 + C_0^2}},
\end{equation}
and we see that in the limit $ma(\tau)\ll C_0$, $\rho(\tau)\simeq 3P(\tau)\simeq N C_0/a^4(\tau)$, while in the limit $C_0\to 0$ or $ma(\tau)\gg C_0$, $\rho(\tau)\simeq N m/a^3(\tau)$ and $P(\tau)\simeq 0$.

In the next section, we combine the results obtained here with Einstein's field equations and discuss the evolution of a universe which is filled with a simple, collisionless gas, with or without a cosmological constant.

\section{The evolution of the universe}

In the previous sections we have derived the one-particle distribution function $f$ and the associated current density and stress-energy tensor for a spatially homogeneous and isotropic, simple, collisionless gas propagating on a fixed FRW background spacetime. The distribution function is characterized by a function $F(C)$ of the single variable $C$ defined in Eq.~(\ref{Def:C}), and once the scale factor $a(\tau)$ is determined, the stress-energy tensor is known. In order to determine $a(\tau)$ in a self-consistent way, we have to solve Einstein's field equations. This is the subject of the present section.

\subsection{Formulation through a one-dimensional mechanical system}

For the FRW metric~(\ref{Eq:FRWMetric}) Einstein's field equations reduce to (see, for instance~\cite{Carroll-Book}) the Friedmann equations
\begin{eqnarray}
\frac{3}{a^2}\left( \frac{da}{d\tau} \right)^2 &=& 8\pi G\rho + \Lambda  - \frac{3k}{a^2},
\label{Eq:Friedmann1}\\
\frac{3}{a}\frac{d^2 a}{d\tau^2} &=& -4\pi G(\rho + 3P) + \Lambda,
\label{Eq:Friedmann2}
\end{eqnarray}
with Newton's constant $G$ and the cosmological constant $\Lambda$. Here, we assume that the energy density $\rho$ and isotropic pressure $P$ are those of a simple, collisionless gas, see Eq.~(\ref{Eq:rhoP}), and hence $P$ and $\rho$ depend themselves on the scale factor $a(\tau)$. Since by construction the expressions for $\rho$ and $P$ given in Eq.~(\ref{Eq:rhoP}) satisfy Eq.~(\ref{Eq:DivTZero}) and since Eq.~(\ref{Eq:Friedmann2}) is a consequence of Eq.~(\ref{Eq:Friedmann1}) and Eq.~(\ref{Eq:DivTZero}), it is sufficient to consider the first Friedmann equation (\ref{Eq:Friedmann1}) in what follows. Once the function $F(C)$ has been chosen, this equation yields a first-order differential equation for the scale factor $a(\tau)$.

The following analysis is based on the simple observation that the first Friedmann equation~(\ref{Eq:Friedmann1}), together with the expression for $\rho$ given in Eq.~(\ref{Eq:rhoP}), describes a mechanical system for the motion of a one-dimensional particle in a potential. In order to describe this system it is convenient to introduce the following dimensionless quantities:
$$
x(t) := \frac{a(\tau)}{a_0},\quad t := H_0\tau,\qquad
f(\zeta) := (m a_0)^3 F(C),\qquad \zeta := C/(m a_0),
$$
where $a_0 = a(\tau = 0)$ is the present value of the scale factor and $H_0 := \left. d\log a(\tau)/d\tau \right|_{\tau=0}$ is the Hubble constant. Note that by definition, $x(0) = \dot{x}(0) = 1$, where here and in the following the dot denotes differentiation with respect to the dimensionless time parameter $t$. In terms of these quantities Eq.~(\ref{Eq:Friedmann1}) can equivalently be written as
\begin{equation}
\dot{x}^2 + V(x) = \Omega_c,
\label{Eq:1DMech}
\end{equation}
with the effective potential
\begin{equation}
V(x) = -\frac{\mu}{x^2}\int\limits_0^\infty f(\zeta)\sqrt{x^2 + \zeta^2}\zeta^2 d\zeta
 - \Omega_\Lambda x^2,
\label{Eq:V}
\end{equation}
where we have defined $\mu := 32\pi^2 G m/(3a_0^3 H_0^2)$ and introduced the usual density parameters $\Omega_\Lambda := \Lambda/(3H_0^2)$ and $\Omega_c := -k/(H_0^2 a_0^2)$ (see, for instance~\cite{Carroll-Book}). Evaluating Eq.~(\ref{Eq:1DMech}) at $t = 0$ yields the familiar constraint
$$
1 = \Omega_M + \Omega_\Lambda + \Omega_c,\qquad
\Omega_M := \mu\int\limits_0^\infty f(\zeta)\sqrt{1 + \zeta^2}\zeta^2 d\zeta
$$
for the density parameters.

Therefore, the dynamics of the scale factor $a(\tau)$ determining the evolution of the universe is equivalent to the dynamics of a one-dimensional point particle with energy $\Omega_c$ propagating in the potential $V(x)$ given in Eq.~(\ref{Eq:V}) whose initial conditions are $x(0) = \dot{x}(0) = 1$. Consequently, it is sufficient to analyze the basic properties of the potential $V(x)$ in order to understand the qualitative features of the evolution of the universe.

As long as $f$ is nonnegative and decays sufficiently fast such that $\Omega_M$ is finite, the effective potential $V(x)$ is well-defined for all $x > 0$. Furthermore, using Lebesgue's dominated convergence theorem, it is not difficult to check that
$$
\lim_{x\to 0} x^2 V(x) = -\mu\int\limits_0^\infty f(\zeta)\zeta^3 d\zeta,
$$
while
$$
\lim_{x\to \infty} x(V(x) + \Omega_\Lambda x^2) 
 = -\mu\int\limits_0^\infty f(\zeta)\zeta^2 d\zeta.
$$
By successive differentiation of $V(x)$ we obtain
\begin{eqnarray}
\frac{dV(x)}{dx} &=& \frac{\mu}{x^3}
 \int\limits_0^\infty f(\zeta) \frac{x^2 + 2\zeta^2}{\sqrt{x^2 + \zeta^2}}\zeta^2 d\zeta
 - 2\Omega_\Lambda x,
\label{Eq:dVdx}\\
\frac{d^2V(x)}{dx^2} &=& -\frac{\mu}{x^4}
 \int\limits_0^\infty f(\zeta) \frac{2x^4 + 9x^2\zeta^2 + 6\zeta^4}{(x^2 + \zeta^2)^{3/2}}\zeta^2 d\zeta
 - 2\Omega_\Lambda.
\label{Eq:ddVddx}
\end{eqnarray}

\subsection{The dynamics for $\Lambda = 0$}

\begin{figure}[ht]
\centerline{\includegraphics[width=10cm,height=7cm]{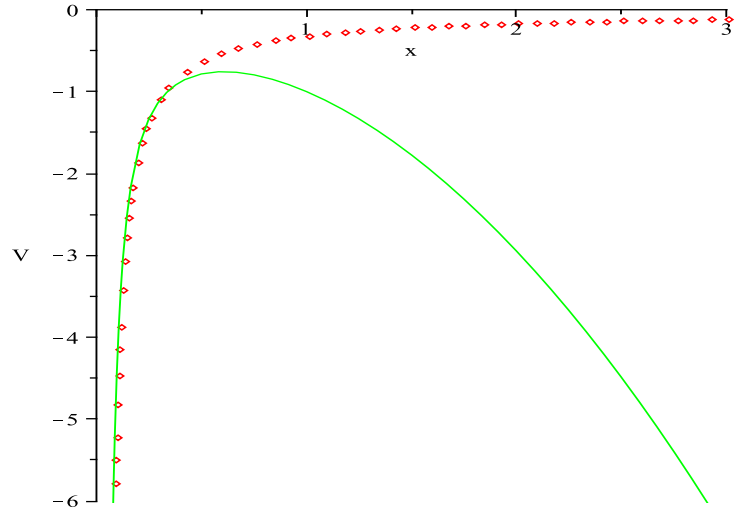}}
\caption{The effective potential $V(x)$ for the case where the distribution function has the form given in Eq.~(\ref{Eq:DeltaF}) with $C_0/(ma_0) = 0.5$. The parameters used are $\Omega_M = 0.3$, $\Omega_\Lambda = 0.0$ (red dotted line) and $\Omega_M = 0.3$, $\Omega_\Lambda = 0.7$ (green solid line).}
\label{Fig:Potential}
\end{figure}

When the cosmological constant is zero ($\Omega_\Lambda = 0$), it follows from Eqs.~(\ref{Eq:V},\ref{Eq:dVdx},\ref{Eq:ddVddx}) that $V(x)$ is a negative, monotonously increasing and concave function with asymptotics $V(x) \simeq -C_3/x^2$ for $x\to 0$ and $V(x) \simeq -C_2/x$ for $x\to \infty$ where $C_2,C_3$ are positive constants. Therefore, the potential looks qualitatively like the function plotted in dotted red in Fig.~\ref{Fig:Potential}.

From this we can immediately reach the following conclusions: For spatially flat ($k=0$) or hyperboloid ($k=-1$) universes, the trajectory $x(t)$ exists for all times in the interval $t\in (-t_s,\infty)$, where
\begin{equation}
t_s = \int\limits_0^1 \frac{dx}{\sqrt{\Omega_c - V(x)}} < \infty
\label{Eq:ts}
\end{equation}
is the age of the universe in units of Hubble time $H_0^{-1}$. The dimensionless scale factor $x(t)$ increases monotonically from $x = 0$ to $x = \infty$, describing a universe which continuously expands from the big bang singularity. The expansion is decelerating since $-dV/dx(x) < 0$. As $\tau\to \infty$, $a(\tau)$ grows like $\tau^{2/3}$ and $\tau$, respectively, depending on whether $k=0$ or $k=-1$, and $\rho$ decays as $1/a^3(\tau)$ and $P$ as $1/a^5(\tau)$ as can be inferred from Eq.~(\ref{Eq:rhoP}).

For a universe with compact spherical sections ($k=1$) the energy of the particle $\Omega_c < 0$ is negative, implying a turning point in the particle's trajectory. Therefore, in this case the universe expands from the big bang singularity, reaches a maximal volume and recollapses to a crunch singularity in a finite amount of proper time.

At the big bang or crunch singularities the scale factor $a(\tau)$ goes to zero and both the energy density $\rho(\tau)$ and the isotropic pressure $P(\tau)$ diverge like $1/a^4(\tau)$ as can be deduced from Eq.~(\ref{Eq:rhoP}).

\subsection{The dynamics for $\Lambda > 0$}

Recent observations suggest that the universe is currently undergoing an accelerated stage of expansion. However, as we have seen above, an accelerated expansion cannot be explained by a collisionless kinetic gas with nonnegative distribution function, since the effective potential $V(x)$ is monotonously increasing in this case. In fact, this can also be seen directly from the second Friedmann equation~(\ref{Eq:Friedmann2}) which shows that any matter field with positive $\rho + 3P$ leads to deceleration.

The simplest manner to explain the current observations is to introduce a cosmological constant $\Lambda$ which is large enough such that $\Lambda > 4\pi G(\rho + 3P)$ for late enough times, inducing an accelerated expansion. For this reason, in what follows we analyze the behavior of the effective potential $V(x)$ given in Eq.~(\ref{Eq:V}) assuming $\Omega_\Lambda > 0$.

When $\Omega_\Lambda > 0$, it still follows from Eqs.~(\ref{Eq:V},\ref{Eq:ddVddx}) that $V(x)$ is negative and concave, and that $V(x) \simeq -C_3/x^2$ for $x\to 0$. However, due to the presence of the cosmological constant, the potential now diverges for $x\to \infty$ since $V(x) \simeq -\Omega_\Lambda x^2$ in this limit. There is a unique maximum $x_*  > 0$ of $V(x)$ which, according to Eq.~(\ref{Eq:dVdx}), is determined by the implicit equation
\begin{equation}
2\Omega_\Lambda x_*^4 
 = \mu\int\limits_0^\infty f(\zeta) \frac{x_*^2 + 2\zeta^2}{\sqrt{x_*^2 + \zeta^2}}\zeta^2 d\zeta.
\label{Eq:VMax}
\end{equation}
Therefore, the effective potential looks qualitatively like the function plotted in solid green in Fig.~\ref{Fig:Potential}. For a spatially flat ($k=0$) or hyperboloid ($k=-1$) universe, for which $\Omega_c \geq 0$, the trajectory $x(t)$ exists again for all times $t\in (-t_s,\infty)$, with $t_s$ given by Eq.~(\ref{Eq:ts}), and the universe continuously expands from the big bang singularity. However, now the expansion is only decelerating until the time $t = t_*$ the maximum of the potential is reached: $x(t_*) = x_*$. After that time, the effects of the cosmological constant become important and the universe starts its phase of accelerated expansion. At late stages the scale factor $a(\tau)$ grows exponentially fast in $\tau$.

For a universe with compact spherical sections ($k=1$), $\Omega_c < 0$ and the qualitative behavior depends on whether or not $\Omega_c$ is larger than the maximum value of $V$. If $\Omega_c > V(x_*)$, the qualitative behavoir of $x(t)$ is the same as in the previous cases $k=0,-1$. However, when $\Omega_c < V(x_*)$, there are two trajectories; one describing a universe that expands from the big bang, reaches a maximal volume and recollapses to a crunch singularity in a finite amount of proper time and one that contracts from $x = \infty$ to a minimal value and starts expanding again to $x = \infty$. If we accept the fact that there was a big bang in the past, the first trajectory is relevant and $x_* > x(0) = 1$. Combining Eqs.~(\ref{Eq:V},\ref{Eq:VMax}) with each other we obtain the estimate
$$
-3\Omega_\Lambda x_*^2 < V(x_*) < -2\Omega_\Lambda x_*^2
$$
for the maximal value of $V$, and consequently $\Omega_c < V(x_*)$ and $x_* > 1$ can only occur if
$$
\Omega_c < -2\Omega_\Lambda.
$$
Recent cosmological observations indicate that the current values of $(\Omega_M,\Omega_\Lambda,\Omega_c)$ lie around $(0.3,0.7,0.0)$, implying that this inequality cannot hold. Furthermore, by evaluating the slope of the potential today we obtain from Eq.~(\ref{Eq:dVdx}),
$$
\frac{dV(1)}{dx} 
 = \mu\int\limits_0^\infty f(\zeta) \frac{1 + 2\zeta^2}{\sqrt{1 + \zeta^2}}\zeta^2 d\zeta
 - 2\Omega_\Lambda \leq 2(\Omega_M - \Omega_\Lambda),
$$
which is negative, indicating that we are currently in the phase of accelerated expansion.

\section{Conclusions}
\label{Sec:Conclusions}

In this work, we have discussed a few basic properties of the Einstein-Liouville system under the assumption that the spacetime metric and the distribution function which describes a collisionless, uncharged, simple, massive gas is homogenous and isotropic. We have defined in detail this homogeneity and isotropicity property of the distribution function, and we have shown that these symmetry requirements and the Liouville equation restrict considerably its dependance upon the phase space variables. This simplification allows one to cast the Einstein-Liouville system (with or without cosmological constant) into a single, ordinary differential equation for the scale factor which can be interpreted as a one-dimensional mechanical system with an effective potential. The analysis of this system shows that irrespectively of the curvature of the spatial sections, any spatially homogeneous and isotropic universe has a singular origin where the curvature, the density and pressure of the gas become unbounded. Close to the singularity, the collisionless gas behaves as a radiation fluid. Furthermore, spatially flat and hyperboloidal universes expand forever, and as we have shown, at late times the kinetic gas behaves much like a pressureless perfect fluid. In this context, we also mention the recent work by Ringstr\"om~\cite{Ringstrom-Book} which proves the nonlinear stability of the spatially flat model with positive cosmological constant in the expanding direction.

It ought to be mentioned that the conclusion of the present work should be viewed with caution. The collisionless nature of the gas may be incompatible with the unbounded growth of the density. One expects that at high densities the mean free path between collisions becomes very small, suggesting that one may not be allowed to neglect collisions between the gas particles. We will return to this question in future work, where the relativistic Boltzmann equation for a homogeneous and isotropic kinetic gas on a FRW background will be analyzed.


\begin{theacknowledgments}
This work was supported in part by a CIC Grant from Universidad Michoacana.
\end{theacknowledgments}

\bibliographystyle{aipproc}
\bibliography{refs_kinetic}

\end{document}